\documentclass[11pt,preprint]{aastex}

\begin{document}

\newcommand{\alex}[1]{({\bf\textsl{#1}})}
\newcommand{\andrei}[1]{({\bf\textsf{#1}})}

\title{Bounds on the Magnetic Fields in the Radiative Zone of the Sun}

\author{Alexander Friedland}
\affil{School of Natural Sciences, Institute for Advanced Study,
Princeton NJ, USA 08540}
\email{alex@ias.edu}

\and

\author{Andrei Gruzinov}
\affil{New York University, 4 Washington Place, New York, NY 10003}
\email{andrei@physics.nyu.edu}

\begin{abstract}
  We discuss bounds on the strength of the magnetic fields that could
  be buried in the radiative zone of the Sun.  The field profiles and
  decay times are computed for all axisymmetric toroidal Ohmic decay
  eigenmodes with lifetimes exceeding the age of the Sun. The
  measurements of the solar oblateness yield a bound $\lesssim 7$ MG
  on the strength of the field. A comparable bound is expected to come
  from the analysis of the splitting of the solar oscillation
  frequencies. The theoretical analysis of the double diffusive
  instability also yields a similar bound. The oblateness measurements
  at their present level of sensitivity are therefore not expected to
  measure a toroidal field contribution.
\end{abstract}


\section{Introduction}

It has been speculated that the solar radiative interior may contain a
buried magnetic field, which does not penetrate into the convective zone
and hence is very difficult to observe. Such a field could have
formed, for instance, during the early stages of the solar evolution,
perhaps as a result of the differential rotation which could have
stretched the lines of a weak primordial seed field
\citep{Dicke79}. Once formed, the large scale field structures would
survive until the present time because of the high conductivity of the
plasma in the radiative zone.

The observation that the lifetime of the large scale field eigenmodes
in the solar interior may exceed the solar age was first made by
Cowling over half a century ago \citep{Cowling1945}.  Cowling's
conclusion was confirmed by other authors, in particular by
\citet{Wrubel52} who computed the lifetimes of the three longest
living poloidal modes and by \citet{BahcallUlrich1971} who computed
the lifetime of the lowest mode using an early solar model.

The motivation for discussing the hypothetical radiative zone (RZ)
field changed with time. The RZ fields, with a characteristic strength
of $10^8$ G, were invoked by several authors
\citep{Bartenwerfer,Chitre} in the 1970's as a possible explanation
for the deficit of solar neutrinos reported by the Homestake
experiment. The fields of comparable strength was postulated by Dicke
\citep{Dicke78,Dicke79,Dicke82} to explain the so-called Princeton
solar oblateness measurements.  Recently, \citet{inevitability} argued
that a relatively weak, but nonzero, RZ poloidal field is necessary to
explain the uniform rotation of the radiation zone. Separately, the
fields of 30 MG strength were proposed as a way to improve the
agreement between the solar model and the measurements of the $^8$B
neutrino flux at the SNO and SuperKamiokande experiments
\citep{Kosovichev02}. In a still separate development, it was found
that the presence of a submegagauss toroidal field in the radiative
zone could explain all presently available solar neutrino data, if
neutrino possesses a relatively large transition magnetic moment
\citep{ourprl}.

%Such a field should be distinguished from the large-scale toroidal
%fields widely believed to exist inside the solar convective
%zone. While the convective zone fields vary with the solar cycle and,
%in particular, reverse direction every 11 years, the hypothetical
%radiation zone fields would be practically constant over this
%timescale. Moreover, while the convective zone fields manifest
%themselves in the wide variety of magnetic phenomena on the surface of
%the Sun, the radiative zone field, if its lines do not penetrate into
%the convective zone, may be very difficult to observe.  

Regardless of the motivation, it is important to establish what
bounds, both observational and theoretical, exist on the strength of
the radiative zone fields. This is the goal of the present paper. Both
Cowling and Wrubel in their analyses assumed a poloidal topology for
the primordial field.  Such field cannot be stronger than a few gauss
at the top of the radiative zone, otherwise it would penetrate in the
convective zone (CZ) and cause a polarity asymmetry between the two
halves of the magnetic cycle \citep{LevyBoyer82,BoyerLevy84,Boruta}.
The absence of the asymmetry places a very strong bound 
($B\lesssim 10^{2}- 10^{3}$ G) on the strength of the poloidal field
allowed in the solar core. At the same time, the field which is
entirely confined to the radiative zone, in particular a toroidal
axisymmetric field, is not subject to this bound, and may in fact be
very difficult to observe.

Before presenting the constraints on the strength of the toroidal
field, in Section~\ref{sect:eigenmodes} we compute the Ohmic decay
eigenmodes for the axisymmetric toroidal field configurations and the
corresponding lifetimes, and give the profiles and lifetimes for all
modes whose lifetimes exceed the solar age. To the best of our
knowledge, such a calculation has not been done previously. The
problem was touched on by \citet{Dicke82}, who presented the profile
of the $l=2$ mode (which is not the longest living mode), without
specifying its lifetime. The discussion of the bounds on the field
strength follows in Section~\ref{sect:constraints}. This is followed
in Section~\ref{sect:theory} by the discussion of theoretical
instabilities that may provide bounds on the strength of the field. In
Section~\ref{sect:discussion} we summarize our conclusions.

\section{Eigenmodes}
\label{sect:eigenmodes}

\begin{figure}
\plottwo{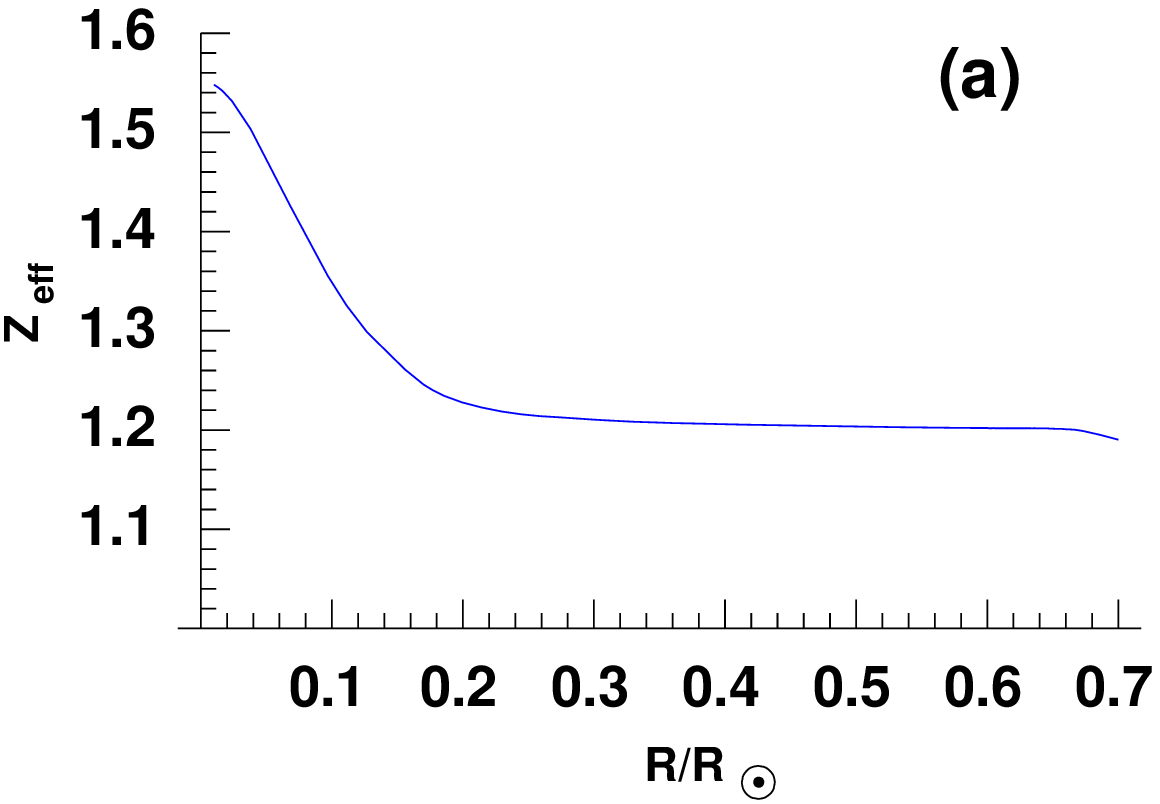}{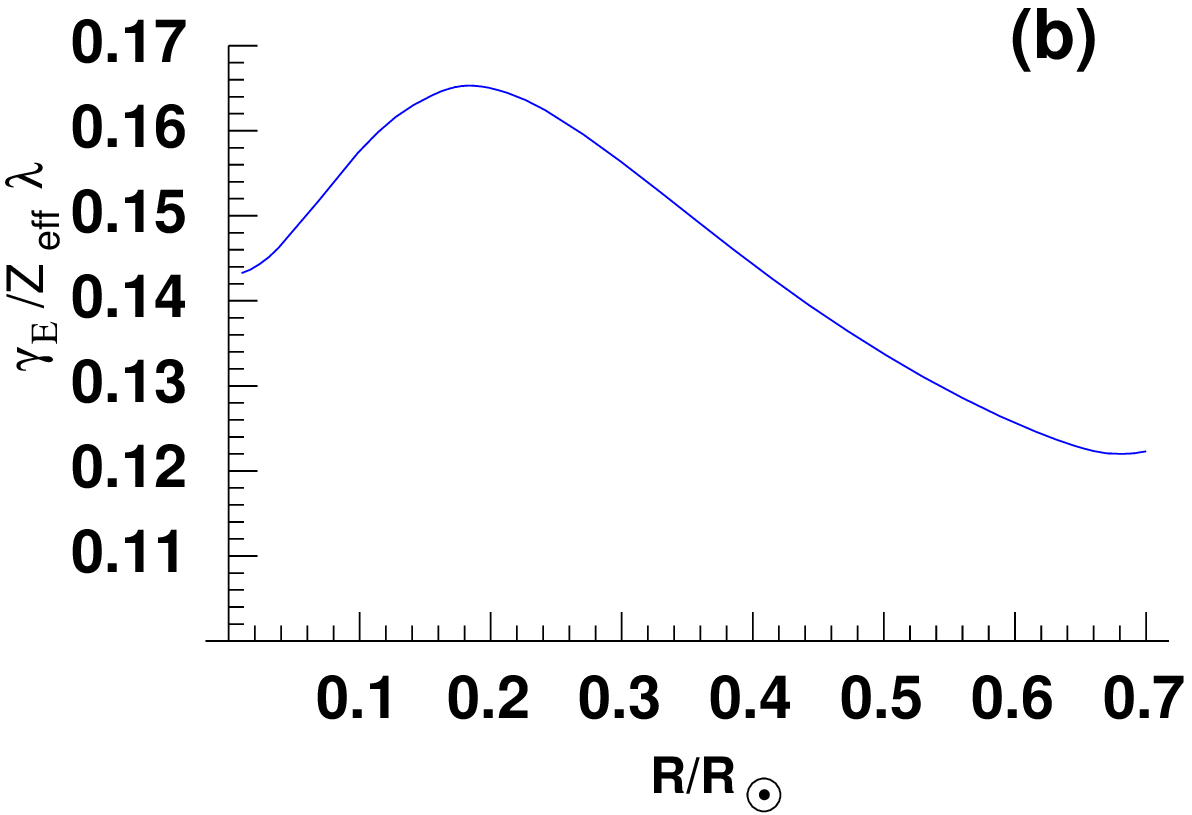}
\caption{The radial dependence of the effective ion charge $Z_{\rm
    eff}$ (left) and the quantity $\gamma_E/Z_{\rm eff}\lambda$
  (right), computed for the BP2000 Sun. \label{fig:gammaEoverZlambda}} 
\end{figure}

We will approximate the electric conductivity in the radiative zone of
the Sun by the Spitzer formula \citep{Spitzerbook}
\begin{equation}
  \label{eq:spitzer}
  \sigma = \frac{2^{5/2}}{\pi^{3/2}}\frac{(k T)^{3/2}}{m_e^{1/2}e^2} 
  \frac{\gamma_E}{Z_{\rm eff} \lambda},
\end{equation}
where $m_e$ and $e$ are the electron's mass and charge, 
$Z_{\rm eff}=\sum_i n_i Z_i^2/\sum_i n_i Z_i$ is the effective ion charge,
$T$ is the temperature, $\gamma_E$ is an order one coefficient
describing the deviation of the plasma from the ideal Lorentz gas, and
$\lambda$ is the Coulomb logarithm.  Numerically, this yields
\begin{equation}
  \label{eq:spitzer_num}
    \sigma = (2.37\times 10^{17} {\rm s}^{-1}) T_6^{3/2} 
  \frac{\gamma_E}{Z_{\rm eff} \lambda},
\end{equation}
where $T_6$ is the temperature in millions of Kelvins. The quantity
$Z_{\rm eff}$ can be computed by using the tables of the solar model;
its dependence on the solar radius for the BP2000 solar model
\citep{BP2000} is shown in Fig.~\ref{fig:gammaEoverZlambda}(a). The
Coulomb logarithm $\lambda$ for $T_6\gtrsim 0.6$ is given by
\begin{equation}
  \label{eq:Lambda}
  \lambda = 2.4 - 1.15 \log_{10} n + 2.3 \log_{10} T_6,
\end{equation}
where the number density
of electrons $n$ is in units of Avogadro number per cm$^3$. Finally,
the coefficient $\gamma_E$ for $Z_{\rm eff}$ in the interval
$1<Z_{\rm eff}<2$ is well described by 
\begin{equation}
  \label{eq:gammaE}
  \gamma_E=0.58 + 0.33 \log_{10} Z_{\rm eff}.
\end{equation}
The radial dependence of the factor $\gamma_E/Z_{\rm eff} \lambda$ in
the Sun is shown in Fig.~\ref{fig:gammaEoverZlambda}(b).

The principal approximation affecting the accuracy of
Eq.~(\ref{eq:spitzer_num}) is using the Spitzer conductivity in the
center of the Sun, where Coulomb logarithm is not large,
$\lambda\approx 2.9$.
%However,
%the resulting small error in the lifetime of the relic field is not
%important.

The corresponding magnetic diffusivity, $\eta=c^2/(4\pi \sigma )$ is
given by
\begin{equation}
  \label{eq:eta}
\eta =(3.0\times 10^{2}{\rm cm}^2{\rm s}^{-1})\frac{Z_{\rm eff}
  \lambda}{\gamma_E} 
T_6^{-3/2}.
\end{equation}
In the convective zone, the effective magnetic diffusivity is very
large because of turbulence.

The Ohmic decay of the field is governed by the magnetic diffusion
equation,
\begin{equation}
  \label{eq:diffusion}
  \partial _t {\bf B}=-\nabla \times (\eta \nabla \times {\bf B}),
\end{equation}
which follows from $\partial _t {\bf B}/c=-\nabla \times{\bf E}$,
${\bf E}= {\bf j}/\sigma$, and $\nabla\times {\bf B}= 4\pi{\bf j}/c$.
Eqs.~(\ref{eq:eta},\ref{eq:diffusion}) can be used to estimate the
minimal size $l_{\rm min}$ of the magnetic field features that 
survive over the age of the Sun, $l_{\rm min}\sim \sqrt{\eta
t_\odot}$, where  $t_\odot=4.6\times 10^9$ years is the age of the 
Sun.  Since $\eta$ varies with the distance from the center, so does
$l_{\rm min}$. At $r=0.6 R_\odot$ one finds  $l_{\rm min}\sim R_\odot/9$,
while at $r=0.25 R_\odot$ $l_{\rm min}\sim R_\odot/20$. Thus, field
modes with the size comparable to the size of the radiative zone are
expected to survive until the present time. 

To make the preceding rough estimate more precise,
we next compute the profiles and the lifetimes of the axisymmetric
toroidal eigenmodes that are localized to the solar RZ.
For an axisymmetric toroidal field
${\bf B}=B(r,\theta, t) {\bf \hat e}_\phi$, Eq.~(\ref{eq:diffusion}) gives
\begin{equation}
\partial _tB=\eta \left( \nabla ^2B-{1\over r^2\sin ^2\theta
    }B\right)~+~\eta 'r^{-1}\partial _r(rB),
\end{equation}
where $'=d/dr$. The eigenmodes have the form
$B=e^{-t/\tau_{n,l}}F_{n,l}(r)P_l^1(\cos\theta)$ where $n=1,2,...$,
$l=1,2,...$. The function $F_ {n,l}$ satisfies the equation
\begin{equation}
\eta \left(F''+2r^{-1}F'-l(l+1)r^{-2}F\right)+\eta '(F'+r^{-1}F)
=-\tau_l^{-1}F.
\end{equation}
The boundary conditions are $F=0$ at $r=0$ and $r=R_{\rm CZ}$.

\begin{figure}
\plottwo{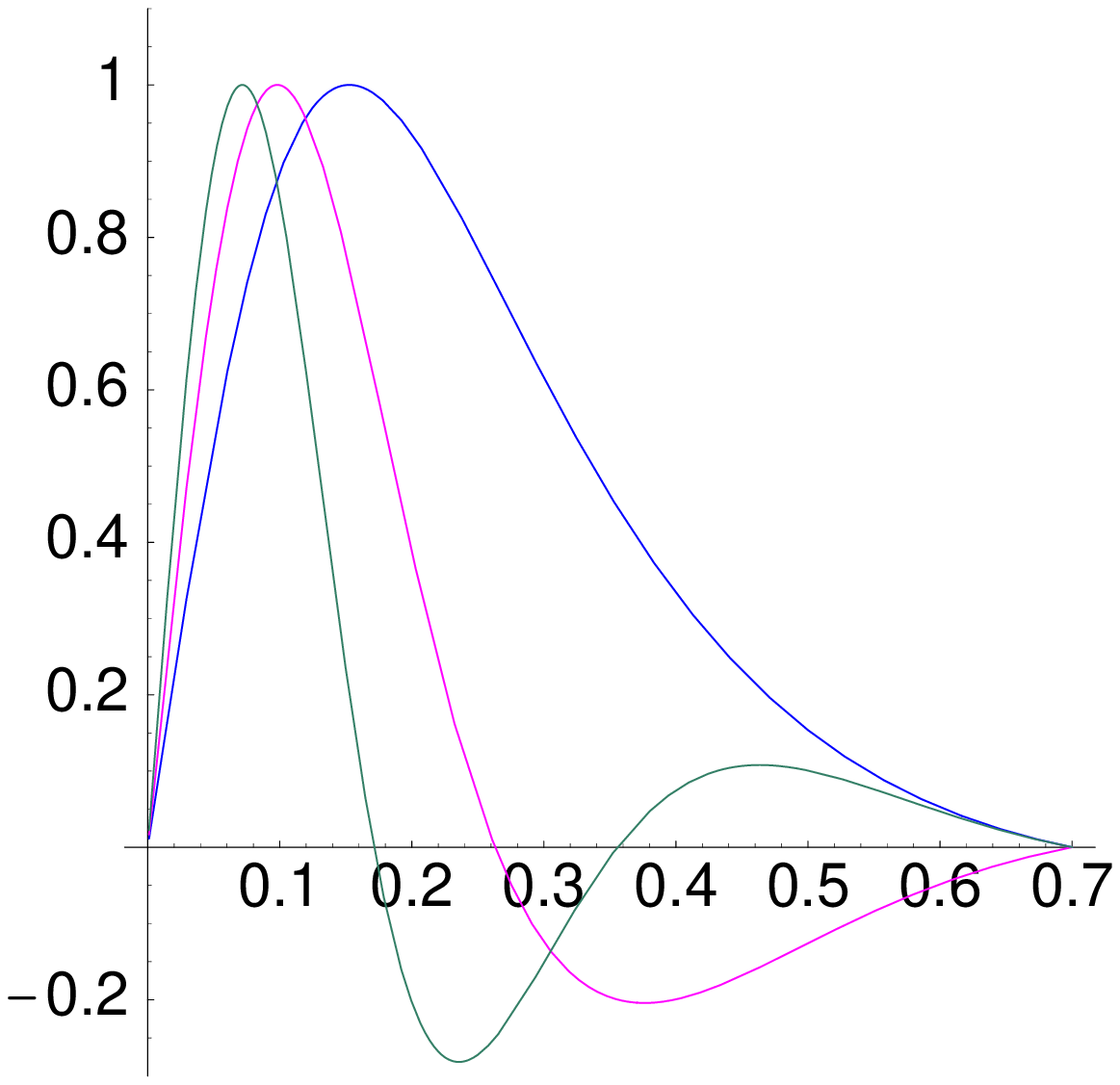}{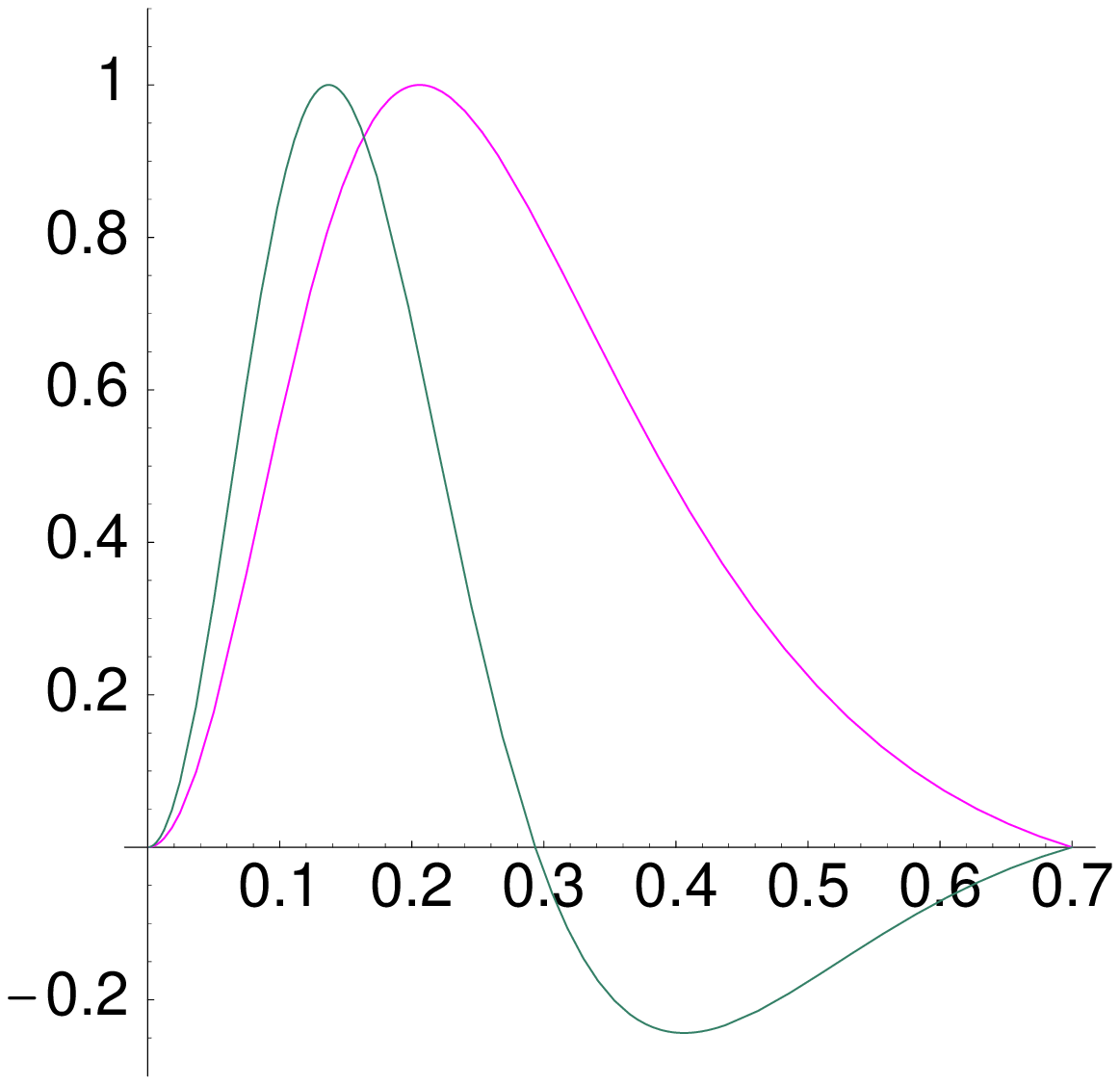}
\caption{The radial dependence $F_{n,l}(r)$ of the  $l=1$
  eigenmodes (left) and $l=2$ eigenmodes (right) whose lifetimes
  exceed the age of the Sun. \label{fig:l1}}
\end{figure}

\begin{figure}
\plottwo{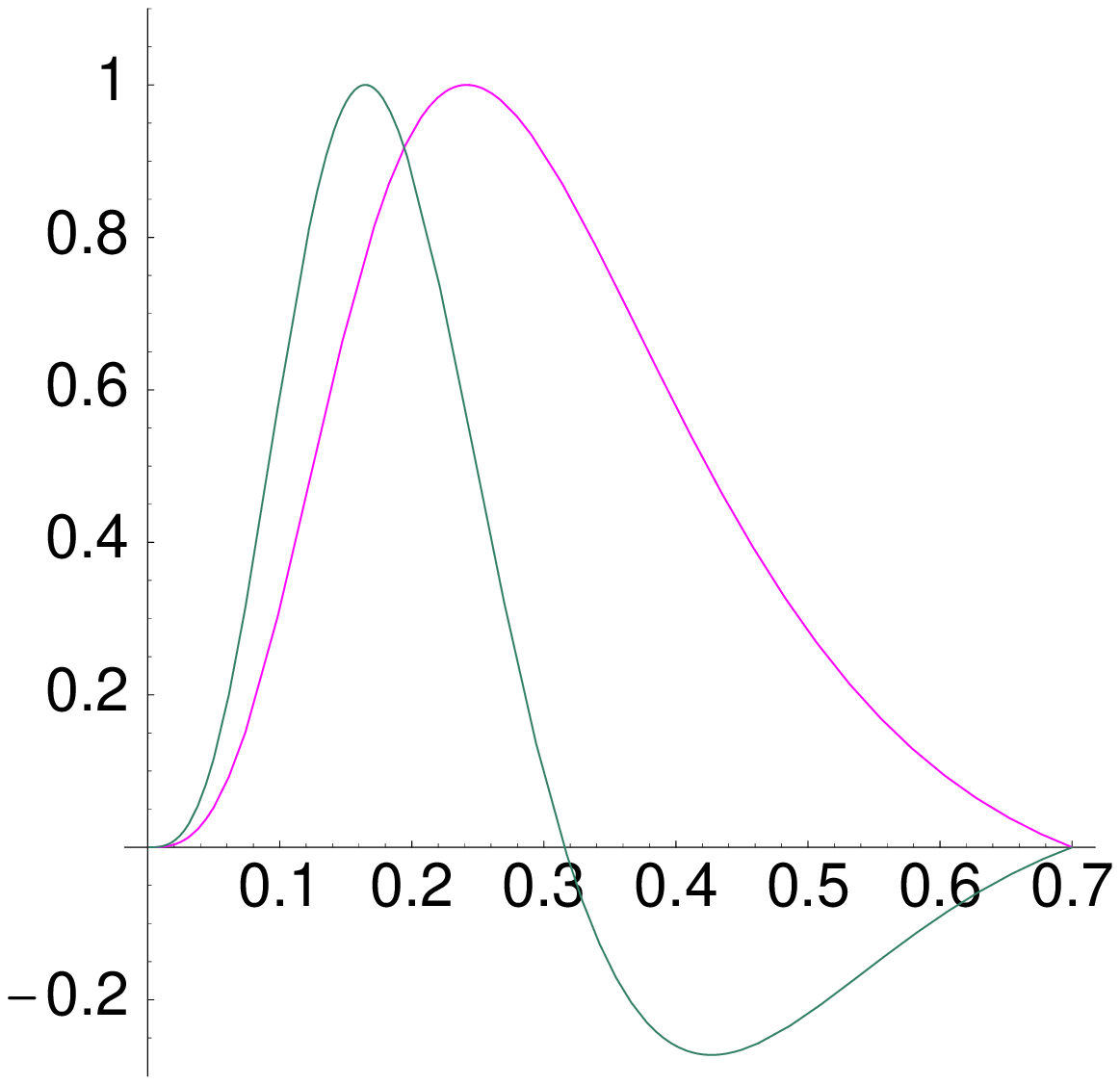}{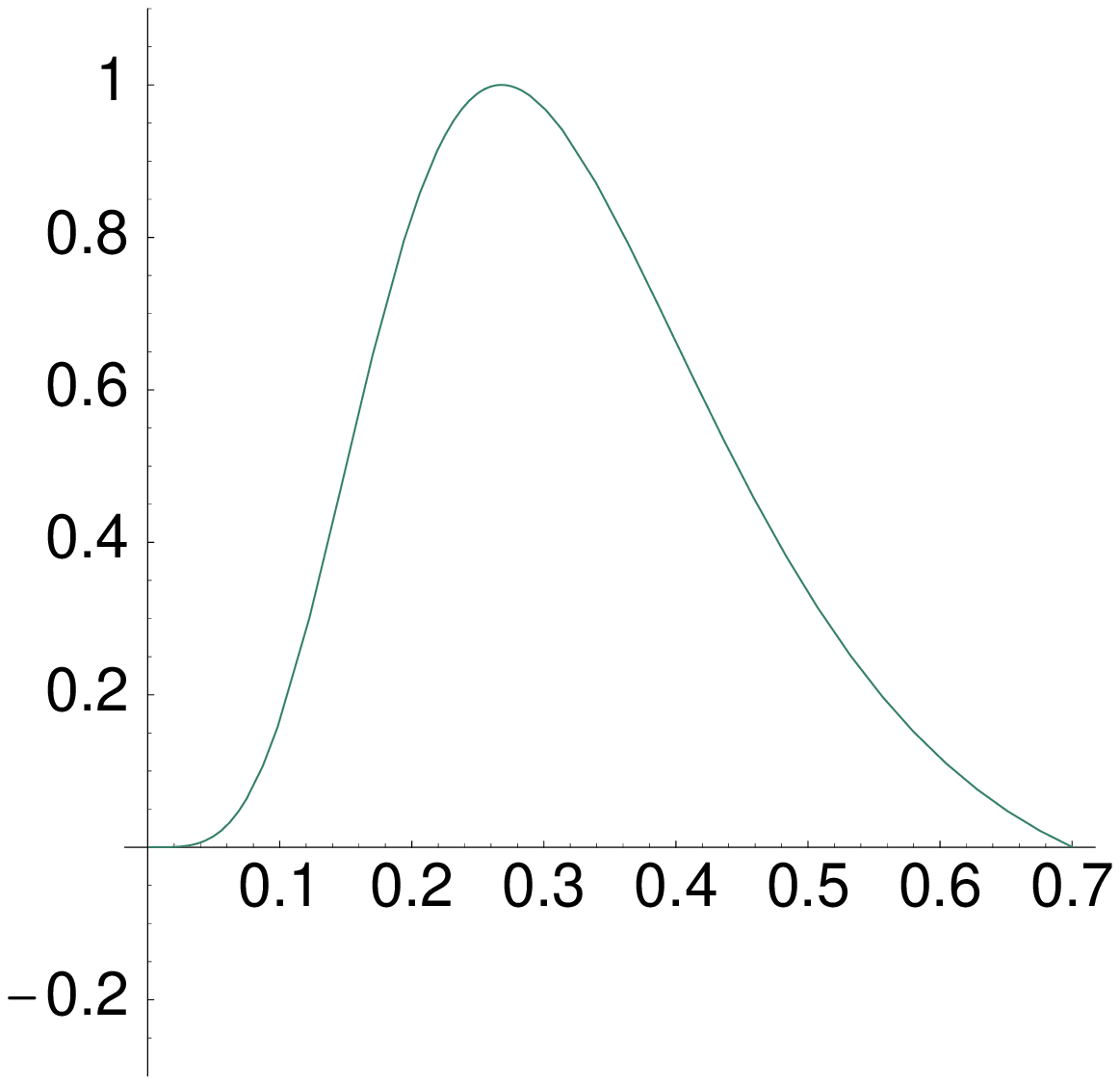}
\caption{The radial dependence $F_{n,l}(r)$ of the $l=3$ (left) and
  $l=4$ (right) eigenmodes whose lifetimes
  exceed the age of the Sun. \label{fig:l2}}
\end{figure}

The lifetimes of various modes, computed for the BP2000 Sun \citep{BP2000}, are
tabulated in Table \ref{table:lifetimes}. The $n=1$, $l=1$ mode has
the longest lifetime,
$\tau_{1,1}= 24{\rm Gyr}$. As can be seen from the table, there are
eight modes whose lifetimes exceed the age of the Sun and three modes
whose lifetimes exceed 10 billion years. Therefore, the toroidal field
in the radiation zone of the Sun can in principle have complex
structure. The profiles $F_{n,l}(r)$ of the eight
longest living modes are shown in Figs.~\ref{fig:l1} and \ref{fig:l2}
and tabulated in Table~\ref{table:Fln}.

Although the preceding calculation does not take into account the
changes of the solar parameters with time, the effect of the solar
evolution on the lifetimes is expected to be small. As a result of
hydrogen burning, as the Sun evolves from $t=0$ to $t=4.7$ Gyr, the
core contracts and becomes hotter, while the outer layers of the RZ
expand and cool.  Since the lifetime depends on the product $l^2
T^{3/2}$, there is partial cancellation, both in the core and in the
outer part of the RZ, between the effects of changing temperature and
changing size. Moreover, the crossover radius, at which the
temperature and the radius distance do not change, is at about $0.15
R_\odot$ \citep{Demarque91}, close to where the maximum of the $n=1$,
$l=1$ mode.

\section{Observational upper bounds on the magnetic field}
\label{sect:constraints}

In this Section we discuss observational upper bounds on the allowed
strength of the magnetic field in the radiation zone. A variety of
independent arguments rule out fields in excess of $\lesssim 10^8$ G.
We briefly summarize the arguments from helioseismological
measurements of the sound speed and from the measurements of the $^8$B
neutrino flux.
%, and from the rise time of the field. 
A much stronger bound on the axisymmetrical fields comes from the
measured oblateness of the Sun. We derive a bound on the strength of
the field specializing to the longest living $n=1$, $l=1$ mode. We also
point out that a comparable bound could be obtained from the
helioseismological measurements of the mode splitting.

\subsection{Bounds from helioseismology and $^8$B neutrino flux}
\label{sect:soundneutrino}

The magnetic field of strength $B$ adds a contribution to pressure
$p_m=B^2/8\pi$, correspondingly decreasing the gas pressure $p_{\rm g}$. 
By estimating the change in the gas temperature as 
$|\delta T/T| \sim p_m/p_{\rm g}$, one obtains a bound on the allowed
strength of the field in the core and in the RZ from the observed flux
of the $^8$B 
neutrinos and the helioseismological measurements of the sound speed.
The two constraints are complimentary, because the $^8$B neutrinos are
produced very close to the center of the Sun ($r<0.09 R_\odot$), where
the accuracy of the helioseismological measurements is relatively poor.

The flux of the $^8$B neutrinos neutrinos is not directly tied to the
solar luminosity and has a strong dependence on the core temperature,
$\phi(^8\mbox{B})\propto T^\beta$, where $\beta\sim 24$
\citep{BahcallUlmer}. Therefore, 
the flux $\phi(^8\mbox{B})$ can be used as a probe of the central
temperature. The flux $\phi(^8\mbox{B})$ has been measured by the SNO
experiment using a neutral current reaction $\nu+d\rightarrow\nu+p+n$
and assuming oscillations into active neutrinos. The measured value,
$5.09\pm 0.64\times 10^{6}$ cm$^{-2}$ s$^{-1}$, is in good agreement
with the predictions of the standard solar model, 
$5.05_{-0.8}^{+1.0}\times 10^{6}$ cm$^{-2}$ s$^{-1}$ \citep{BP2000}.
Combining the above ingredients, we obtain an order of magnitude bound
on the strength of the fields in the core,
\begin{equation}
  \label{eq:neutrinobound}
  B\lesssim p_{\rm g}^{1/2}
\left(\frac{\delta\phi(^8\mbox{B})}{\phi(^8\mbox{B})}\right)^{1/2}  
  \sim 2\times 10^8 \mbox{G}.
\end{equation}
A similar order of magnitude bound follows from the measurements of
the sound speed. Taking the uncertainty on the sound speed to be
$\delta c_{\rm s}/c_{\rm s} \sim 10^{-3}$ we find for $r=0.2 R_\odot$,
\begin{equation}
  \label{eq:soundbound}
  B\lesssim (8\pi p_{\rm g})^{1/2}
\left(\frac{\delta c_{\rm s}}{c_{\rm s}}\right)^{1/2}  
  \sim 0.4\times 10^8 \mbox{G}.
\end{equation}

To obtain more accurate bounds in both cases requires incorporating
magnetic fields in the solar model calculations. Recent analysis by 
\citet{Kosovichev02} finds the limit $B\lesssim 3\times 10^7$ G from
the measurements of the sound speed.

\subsection{Constraints from Solar Oblateness}

It is well known that the rotation distorts the shape of the Sun, making
it oblate. The theoretical value of the solar rotational oblateness
$\epsilon \equiv 1-R_{\rm pole}/R_{\rm equator}$ is $\epsilon _{\rm rot}=9
\times 10^{-6}$ \citep{GodierRozelot}. The magnetic fields will further
distort the equilibrium figure of the Sun. 
If we take the recently measured oblateness of the Sun to be
$(10\pm 3)\times 10^{-6}$ \citep{GodierRozelot}, we must require that the
contributions to the oblateness from the magnetic fields
not exceed the purely rotational oblateness by over $\sim 30$\%,
which translates into a bound on $B_{\rm max}$.

The distortion is obtained from the equilibrium equation
\begin{equation}
\vec\nabla (p+p_m) + {2p_m\over R}\hat{R} + \rho \vec\nabla \phi =0.
\end{equation}
Here $p$ is the gas pressure, $p_m=B^2/(8\pi )$ is the pressure
of the toroidal magnetic field, $R=r\sin \theta$ is the
cylindrical radius, $\hat{R}$ is the unit vector along
cylindrical radius, $\rho$ is the density, $\phi$ is the
gravitational potential. We can exclude $p$ by taking the curl,
and then linearize, that is leave only first order terms in $p_m$:
\begin{equation}
\phi _0'\partial _{\theta }\delta \rho - \rho _0'\partial
_{\theta }\delta \phi = 2\cot \theta \partial _rp_m-2r^{-1}
\partial _{\theta }p_m.
\end{equation}
Here $\delta \rho$ and $\delta \phi$ are density and potential
perturbations due to magnetic pressure, $\rho _0$ and $\phi _0$
are the unperturbed density and potential of the Sun, and prime
is the derivative with respect to $r$.

We now specialize to the longest living mode, $p_m(r, \theta )=B_{\rm
  max}^2 [F_{1,1}(r)]^2\sin^2 \theta/8\pi$, with $F_{1,1}(r)$ tabulated in
Table~\ref{table:Fln}. Then $\delta \rho$ and $\delta \phi$ are
$\propto P_2(\cos \theta )$ and, using the Poisson equation $\nabla^2
\phi = 4\pi G \rho$, we obtain
\begin{equation}
(4\pi G)^{-1}\phi _0'(\delta \phi ''+2r^{-1}\delta\phi '-6r^{-2}\delta\phi) -
\rho _0'\delta \phi = -(4/3) B_{\rm max}^2 F_{1,1}(F_{1,1}'-r^{-1} F_{1,1}).
\end{equation}
This equation was solved numerically, and the calculated
oblateness of the Sun was found to be
\begin{equation}
\epsilon_{\rm magn} =-6.4\times 10^{-8} \left( {B_{\rm max}\over 1{\rm
MG}} \right)^2.
\end{equation}

By requiring that $\epsilon_{\rm magn}\lesssim 3\times 10^{-6}$,
we get an upper bound
\begin{equation}
B_{\rm max}\lesssim 7 {\rm MG}.
\end{equation}
Thus the bound from oblateness is much stronger then the ones
considered earlier. 

The effect of the 12 MG field would be to completely cancel out the
solar rotational oblateness, making the Sun into a perfect sphere. The
fields of the strength $3\times 10^7$ G, that were invoked by
\citet{Kosovichev02} as a possible way to improve the agreement
between the observed and predicted $^8$B fluxes, would make the figure
of the Sun strongly prolate and hence are not allowed.

\subsection{Splitting of Solar Oscillation Frequencies}

The observations of the splitting of solar oscillation frequencies can
provide another way to constrain the strength of the magnetic field.
\citet{AntiaChitreThompson} analyzed the observed splittings and
derived an upper bound of $B<0.3$MG for a toroidal field of radial
extent $d=0.04R_{\odot }$ near the base of the convective zone at
$r=0.7R_{\odot }$. We can write their upper bound in a form \footnote{
  This is a much stronger upper bound than the upper bound $B^2/(8\pi
  p)<10^{-3}$, considered in Sect.~\ref{sect:soundneutrino}.}
\begin{equation}
{d\over r}{B^2\over 8\pi p(r)}~<3~\times 10^{-6}.
\end{equation}
If we assume that the error bars on the measured splitting
coefficients do not increase too much as we move deeper into the Sun,
we obtain an upper bound on the strength of the lowest field mode that
is comparable to the one derived from oblateness.  To make this bound
more precise requires repeating the full analysis of
\citet{AntiaChitreThompson} with the magnetic field profile
composed of the modes found in Sect.~\ref{sect:eigenmodes}.

\section{Theoretical Considerations}
\label{sect:theory}

In this section we describe theoretical bounds on the magnitude of the
radiative zone magnetic field. The bounds are based on the observation that 
a sufficiently strong field will rise to the surface as a result of
thermal diffusion in the solar plasma. We first consider the
simplified scenario in which the field rises as a whole and then
address the possibility that the field may fragment and the fragmented
flux tubes rise to the surface. The latter analysis gives a
particularly stringent bound on the field strength. We also comment on
the role of the ideal magnetohydrodynamic instability that may exist
for the field configuration in question.

\subsection{Constraint from Magnetic Buoyancy}
\label{sect:buoyancy}

A robust bound on the strength of the primordial magnetic field
follows from the consideration of the rise time in the stably
stratified RZ \citep{Parker74,Parkerbook}. The essence of the effect
is that a flux tube in hydrostatic equilibrium in a stably stratified
medium has a lower temperature than the temperature of the surrounding
medium. The resulting flow of heat into the tube causes it to rise to
assume a new equilibrium position. The rise velocity $u$ thus depends
on the field strength (through temperature gradient) and the heat
conduction properties of the plasma,
\begin{equation}
  \label{eq:parkerrise}
  u \simeq \frac{1}{\delta}\frac{B^2}{8\pi p}
  \left(\frac{\lambda}{R}\right)^2 \frac{I}{p}.
\end{equation}
Here $p$ is the equilibrium pressure, $I=L_\odot/4\pi r^2$ is the heat
flux, $R$ is the radius of the flux tube, $\lambda\equiv d\ln T/d r$
is the temperature scale height of the medium, and 
$\delta\equiv (d\ln n/d\ln T)-1/(\gamma-1)$ is a parameter determining
the stability of the medium (for an adiabatic medium $\delta=0$).

Numerically, the rise time from $r=0.2 R_\odot$ to the top of the RZ is  
\begin{equation}
  \label{eq:risetime}
  t_{\rm rise} \sim 3\times 10^7 {\rm yr}\left(\frac{B^2}{8\pi p}\right)^{-1},
\end{equation}
so that the field $B\sim 10^8$ G would rise in the time comparable to
the age of the Sun. For the field $B\sim 0.5\times 10^6$ G considered
in \citet{ourprl} the
distance traveled over the age of the Sun is only about 8 km and the
effect can be safely ignored.

The above mechanism can be used to make an argument that the
primordial RZ field \emph{could not have formed} with the strength of
$10^8$ G.  This is possible using the abundance of Beryllium at the
solar surface. Be is destroyed at $T=3\times 10^6$ K, which is the 
temperature at $r=0.61 R_\odot$. Since Beryllium observed at the solar
surface is not noticeably depleted, the material at $r\sim 0.7 R_\odot$ 
and the material at $r\sim 0.6 R_\odot$ have not been mixed during the
lifetime of the Sun (except for a brief 
period during the pre-main sequence convection stage). The rise of the $10^8$ G
field would have forced such mixing, and therefore could not have happened.

\subsection{Double Diffusive Instability}

The preceding analysis showed that the rise velocity of a flux tube is
inversely proportional to the diameter of the tube squared. The simple
reason for this is that  small flux tubes are heated faster
than large ones. It is therefore conceivable that instead
of rising as a whole, the magnetic field might fragment into small
flux tubes, which would rise to the surface faster. This instability
is counteracted by the diffusion of the magnetic field lines.
Superficially, it may appear that, since the heat diffusivity of the
solar plasma,
\begin{equation}
  \label{eq:chi}
  \chi=\frac{16\sigma T^3}{3 \kappa \rho^2 c_v},
\end{equation}
is some four orders of magnitude greater than the magnetic diffusivity
$\eta$, given in Eq.~(\ref{eq:eta}), any field configuration will be
unstable. However, a more careful consideration of the effect shows
that the relevant
parameter controlling the instability is the combination 
$(\chi/\eta) (p_{\rm m}/p_{\rm g})$. The stability analysis was
carried out by \citet{Schubert}, who found that for the field
configuration to be stable, it is necessary that 
\begin{equation}
  \label{eq:schubert}
  \frac{\partial \ln(p \rho^{-\gamma})}{\partial r} >
  -\frac{\chi}{\eta} \frac{B^2}{4\pi p_{\rm g}} 
  \frac{\partial \ln(B_\phi/\rho r)}{\partial r}.
\end{equation}
Substituting in this condition the profile of the longest living
toroidal eigenmode ($l=1$, $n=1$), we find that if the normalization
of the field, $B_{\rm max}$, exceeds $\sim 2.1$ MG, the field
configuration becomes unstable at $r \sim 0.6 R_\odot$. Therefore,
this argument provides the most stringent bound on the toroidal
magnetic field.

\subsection{Ideal magnetohydrodynamic instability}

An arbitrarily weak purely toroidal magnetic field is unstable to
ideal magnetohydrodynamic (MHD) perturbations 
\citep{Tayler1973,Goossens1978,Goossens1980}.
However, one can show
\citep{preparation} that the perturbation considered by
\citet{Goossens1978} is stabilized by a poloidal field $B_p$ which is
much weaker than the destabilizing toroidal field $B_t$ 
($B_p^2/8\pi p\lesssim (B_t^2/8\pi p)^2$).

The general problem of ideal MHD stability of toroidal plus poloidal
magnetic field in spherical geometry with stable stratification is
unsolved. One also notes that toroidal fields satisfying observational
upper bounds are very weak ($B_t^2/(8\pi p)\lesssim 3\times 10^{-6}$),
and the unstable perturbations have a very small wavelength
\citep{Goossens1978}. Then even a trace of differential rotation in
the RZ can affect stability in an unknown manner. It is therefore
difficult to extract a concrete bound on the strength of the magnetic
field from the analysis of the MHD instability with the present level
of understanding of the problem.

\section{Discussion}
\label{sect:discussion}

We have discussed a variety of constraints on the allowed strength
of the magnetic fields in the solar radiative zone.  We have shown
that the strongest constraints come from the measurements of the solar
oblateness, from the splitting of the oscillation frequencies, and
from the theoretical analysis of the double diffusive instability of
the field. All three constraints point to maximal field values of
less than a few magegauss. That very similar bounds emerge from the
analysis of such physically different phenomena is in itself quite
remarkable.  Alternatively, our results show that magnetic field that
would measurably distort the figure of the Sun would cause an
instability in the Sun, and hence is not allowed. This explains why
the oblateness measurements at their present level of sensitivity have
not seen a toroidal field contribution.

Our analysis excludes the possibility of perturbing the solar model
with the fields of the order of several tens of megagauss that have been
invoked to obtain a better agreement with the solar neutrino data
\citep{Kosovichev02}.

The bounds obtained here have a direct implication for the analysis of
the neutrino survival probability discussed in \citet{ourprl}. The
survival probability depends on the value of the product of the
neutrino transition moment, $\mu$, and the magnetic field
normalization, $B_{\rm max}$. The value of the product, which was
found to give a good fit to the solar neutrino data, is 
$\mu B_{\rm max} \sim 0.5\times 10^{-11}$ MG$\times\mu_B$ ($\mu_B$ is the Bohr
magneton). The upper bound on the magnetic field strength therefore
yields a \emph{lower} bound on the size of the transition moment that
is necessary to fit the solar neutrino data. The
bound from oblateness yields $\mu \gtrsim 7\times 10^{-13}\mu_B$ , while
the limit from the double diffusive instability gives 
$\mu \gtrsim 2\times 10^{-12}\mu_B$.

Among the theoretical issues that require further study is the problem
of ideal MHD stability described in Section~\ref{sect:discussion} and
the possibility that the symmetry axis of the magnetic
field may become tilted as a result of the dissipative motions of the
plasma \citep{Spitzertilt,Mestel1981}.  
%This is an extremely
%interesting possibility which, however, requires a very complicated
%analysis. At present, the problem remains unsolved. 

In summary, the bounds on the strength of the hypothetical magnetic
fields in the radiative zone of the Sun described here may be useful
for searches for such fields and also for the discussions of their
physical implications.

\acknowledgements

We thank John Bahcall for valuable discussions and his encouragements
during the course of this work. We are also grateful to Axel
Brandenburg, Sarbani Basu, and Peter Goldreich for useful
discussions. A. F. was supported by the W.~M.~Keck Foundation.

\clearpage

%\clearpage

%\clearpage

%\begin{figure}
%\plotone{l3modes.eps}
%\caption{The radial dependence $F_{l,n}(r)$ of the $l=3$ eigenmodes. \label{fig:l3}}
%\end{figure}

\clearpage

\begin{table}
\begin{center}
\begin{tabular}{c|rrrrr}
 & 1 & 2 & 3 & 4 & 5 \\
\tableline
1 &      24.3&12.8&7.7&5.2&3.7 \\
2 &      10.0&6.6&4.6&3.4&2.6 \\
3 &      5.3&4.0&3.0&2.4&1.9 \\
4 &      3.3&2.6&2.1&1.7&1.4 \\
5 &      2.2&1.8&1.5&1.3&1.1 
\end{tabular}
\end{center}
\caption{The lifetimes of the field eigenmodes in billions of years as
  a function of $n$ (vertical) and $l$ (horizontal).}
\label{table:lifetimes}
\end{table}

\clearpage

\begin{deluxetable}{c|rrrrrrrr}
\tabletypesize{\scriptsize}
\tablecaption{The radial dependence $F_{l,n}(r)$ of the eigenmodes
  whose lifetimes exceed the age of the Sun.
  \label{table:Fln}}
\tablewidth{0pt}
\tablehead{
\colhead{r} & \colhead{$F_{1,1}$}   & \colhead{$F_{2,1}$}   &
\colhead{$F_{3,1}$}  &  \colhead{$F_{1,2}$}  &
\colhead{$F_{2,2}$} & \colhead{$F_{1,3}$}  & \colhead{$F_{2,3}$} &
\colhead{$F_{1,4}$}}
\startdata      
0.01 & 0.1130 &  0.1669 &  0.2239 & 0.0077 &  0.0147 & 0.0005 &  0.0011 & 0.0000 \\
0.03 & 0.3326 &  0.4809 &  0.6250 & 0.0672 &  0.1271 & 0.0120 &  0.0277 & 0.0020 \\
0.05 & 0.5325 &  0.7378 &  0.8978 & 0.1772 &  0.3228 & 0.0523 &  0.1164 & 0.0146 \\
0.07 & 0.7019 &  0.9118 &  0.9994 & 0.3217 &  0.5537 & 0.1306 &  0.2759 & 0.0501 \\
0.09 & 0.8341 &  0.9929 &  0.9345 & 0.4810 &  0.7673 & 0.2455 &  0.4839 & 0.1184 \\
0.11 & 0.9265 &  0.9862 &  0.7453 & 0.6363 &  0.9219 & 0.3863 &  0.6978 & 0.2216 \\
0.13 & 0.9804 &  0.9081 &  0.4923 & 0.7723 &  0.9948 & 0.5372 &  0.8728 & 0.3528 \\
0.15 & 0.9997 &  0.7809 &  0.2339 & 0.8790 &  0.9834 & 0.6816 &  0.9769 & 0.4987 \\
0.17 & 0.9902 &  0.6270 &  0.0129 & 0.9519 &  0.9006 & 0.8061 &  0.9972 & 0.6435 \\
0.19 & 0.9581 &  0.4661 & -0.1477 & 0.9910 &  0.7671 & 0.9017 &  0.9390 & 0.7727 \\
0.21 & 0.9096 &  0.3128 & -0.2429 & 0.9995 &  0.6062 & 0.9646 &  0.8200 & 0.8761 \\
0.23 & 0.8504 &  0.1766 & -0.2799 & 0.9824 &  0.4381 & 0.9955 &  0.6631 & 0.9486 \\
0.25 & 0.7850 &  0.0623 & -0.2722 & 0.9456 &  0.2782 & 0.9977 &  0.4907 & 0.9890 \\
0.27 & 0.7171 & -0.0288 & -0.2348 & 0.8944 &  0.1365 & 0.9760 &  0.3206 & 0.9999 \\
0.29 & 0.6493 & -0.0976 & -0.1810 & 0.8339 &  0.0179 & 0.9358 &  0.1655 & 0.9852 \\
0.31 & 0.5834 & -0.1465 & -0.1216 & 0.7679 & -0.0759 & 0.8823 &  0.0329 & 0.9500 \\
0.33 & 0.5207 & -0.1785 & -0.0640 & 0.6997 & -0.1460 & 0.8198 & -0.0742 & 0.8995 \\
0.35 & 0.4618 & -0.1967 & -0.0128 & 0.6316 & -0.1946 & 0.7521 & -0.1556 & 0.8383 \\
0.37 & 0.4072 & -0.2038 &  0.0295 & 0.5652 & -0.2250 & 0.6823 & -0.2130 & 0.7703 \\
0.39 & 0.3570 & -0.2026 &  0.0621 & 0.5016 & -0.2402 & 0.6125 & -0.2493 & 0.6990 \\
0.41 & 0.3111 & -0.1951 &  0.0852 & 0.4417 & -0.2434 & 0.5444 & -0.2679 & 0.6268 \\
0.43 & 0.2695 & -0.1832 &  0.0996 & 0.3859 & -0.2375 & 0.4794 & -0.2723 & 0.5559 \\
0.45 & 0.2318 & -0.1685 &  0.1067 & 0.3344 & -0.2248 & 0.4180 & -0.2655 & 0.4875 \\
0.47 & 0.1979 & -0.1520 &  0.1076 & 0.2873 & -0.2074 & 0.3610 & -0.2507 & 0.4229 \\
0.49 & 0.1676 & -0.1347 &  0.1039 & 0.2445 & -0.1871 & 0.3084 & -0.2301 & 0.3627 \\
0.51 & 0.1406 & -0.1173 &  0.0967 & 0.2058 & -0.1653 & 0.2605 & -0.2061 & 0.3073 \\
0.53 & 0.1165 & -0.1003 &  0.0872 & 0.1712 & -0.1430 & 0.2172 & -0.1801 & 0.2567 \\
0.55 & 0.0953 & -0.0840 &  0.0762 & 0.1403 & -0.1209 & 0.1783 & -0.1536 & 0.2111 \\
0.57 & 0.0765 & -0.0689 &  0.0645 & 0.1128 & -0.0998 & 0.1436 & -0.1275 & 0.1702 \\
0.59 & 0.0599 & -0.0549 &  0.0527 & 0.0886 & -0.0799 & 0.1128 & -0.1026 & 0.1339 \\
0.61 & 0.0454 & -0.0421 &  0.0412 & 0.0672 & -0.0616 & 0.0856 & -0.0794 & 0.1017 \\
0.63 & 0.0327 & -0.0306 &  0.0304 & 0.0484 & -0.0449 & 0.0617 & -0.0580 & 0.0734 \\
0.65 & 0.0216 & -0.0204 &  0.0204 & 0.0320 & -0.0299 & 0.0408 & -0.0388 & 0.0487 \\
0.67 & 0.0120 & -0.0113 &  0.0114 & 0.0178 & -0.0167 & 0.0227 & -0.0217 & 0.0272 \\
0.69 & 0.0037 & -0.0035 &  0.0035 & 0.0055 & -0.0051 & 0.0070 & -0.0067 & 0.0087 
\enddata

%% Text for table notes should follow after the \enddata but before
%% the \end{deluxetable}. Make sure there is at least one \tablenotemark
%% in the table for each \tablenotetext.

\end{deluxetable}

\end{document}